\documentclass[english,notitlepage,nofootinbib,reprint]{revtex4-1}
\usepackage[T1]{fontenc}
\usepackage[latin9]{inputenc}
\usepackage{color}
\usepackage{amsmath}
\usepackage{graphicx}
\usepackage{esint}

\makeatletter
\usepackage{slashed}
\usepackage[percent]{overpic}

\definecolor{very_gray}{rgb}{0.3,0.3,0.3}

\makeatother

\usepackage{babel}
\begin{document}

\title{Tensor and scalar interactions of neutrinos may lead to \\
observable neutrino magnetic moments}

\author{Xun-Jie Xu}

\affiliation{\textcolor{black}{Max-Planck-Institut f\"ur Kernphysik, Postfach 103980, D-69029
Heidelberg, Germany}.}

\date{\today}
\begin{abstract}
Recently more generalized four-fermion interactions of neutrinos
such as tensor and scalar interactions (TSIs) have been extensively
studied in response to forthcoming precision measurements of neutrino
interactions. In this letter, we show that due to the chirality-flipping
nature, at the 1-loop level TSIs typically generate much larger ($10^{7}\sim10^{10}$)
neutrino magnetic moments ($\nu$MMs)  than the vector case. For
some cases, the large $\nu$MMs generated by TSIs may reach or exceed
the known bounds, which implies potentially important interplay between
probing TSIs and searching for $\nu$MMs in current and future neutrino
experiments.
\end{abstract}
\maketitle

\section{Introduction}

As neutrino experiments are entering the precision era, searching
for new neutrino interactions beyond the Standard Model (BSM) is of
increasing importance. In the near future, experiments of coherent
neutrino-nucleus scattering\footnote{First observed by the recent COHERENT experiment~\cite{Akimov:2017ade}.
The future experiments include  CONUS \cite{CONUStalk}, $\nu$-cleus \cite{Strauss:2017cuu}, CONNIE \cite{Aguilar-Arevalo:2016khx}, MINER \cite{Agnolet:2016zir}, etc.} and other types of neutrino scattering\footnote{E.g., neutrino scattering at the near detectors \cite{Wang:2017tmk,Bian:2017axs,Falkowski:2018dmy,deNiverville:2018dbu,Bischer:2018zcz,Bakhti:2018avv,Ballett:2018uuc}
of long baseline experiments, or at IsoDAR \cite{Conrad:2013sqa,Abs:2015tbh},
LZ \cite{Coloma:2014hka,Akerib:2015cja}, etc. }, will reach unprecedented sensitivity to various types of BSM neutrino
interactions. 

Among various BSM interactions considered for neutrinos, the so-called
Non-Standard Interactions (NSIs, see reviews \cite{Davidson:2003ha,Ohlsson:2012kf,Farzan:2017xzy,Esteban:2018ppq}),
which couple neutrinos ($\nu$) to other fermions ($\psi$) by the
flavor-changing effective operators $\overline{\nu_{\alpha}}\gamma_{\mu}\nu_{\beta}\overline{\psi}\gamma^{\mu}\psi$,
have been extensively studied due to their rich phenomenology in neutrino
oscillation. In addition to NSIs which are of the vector form (i.e.
containing $\gamma_{\mu}$ between $\overline{\nu}$ and $\nu$),
recently there has been rising interest in more general interactions
\cite{Healey:2013vka,Sevda:2016otj,Lindner:2016wff,Heurtier:2016otg,Rodejohann:2017vup,Kosmas:2017tsq,Magill:2017mps,Farzan:2018gtr,Yang:2018yvk,AristizabalSierra:2018eqm,Bischer:2018zcz,Brdar:2018qqj,Blaut:2018fis}
 of scalar or tensor forms with the $\gamma_{\mu}$ replaced by $\mathbf{1}$
or $\sigma_{\mu\nu}$ respectively\footnote{More generally, one can have additional $\gamma_{5}$'s attached,
which would form pseudoscalar, axial vector and CP-violating tensor
interactions. Hereafter, as a simplified terminology, we will refer
to them as scalar, vector and tensor interactions likewise.}. From the theoretical point of view, the scalar or tensor interactions
are as well motivated as the NSI, since they can all originate from
integrating out some BSM bosons\footnote{Integrating out a vector boson may give rise to NSI while integrating
out a charged scalar boson may lead to both scalar and tensor interactions\,---\,exemplified
later in Sec.~\ref{sec:UV}. }.

In this letter, we would like to point out that the scalar or tensor
interactions of neutrinos may lead to much larger neutrino magnetic
moments ($\nu$MMs) than the vector interactions. For the vector
case, the loop-generated $\nu$MM is proportional to the neutrino
mass and thus highly suppressed \cite{Petcov:1976ff,Marciano:1977wx,Lee:1977tib,Fujikawa:1980yx,Pal:1981rm,Shrock:1982sc}.
However, for scalar or tensor interactions, due to their chirality-flipping
feature as will be explained later, it is proportional to the mass
of $\psi$ \footnote{The idea of obtaining large $\nu$MMs by avoiding it from being proportional
to a neutrino mass  is not new and has been discussed widely in the
literature, see the review \cite{Giunti:2014ixa} and references therein.
For further discussions, see  Sec.~\ref{sec:Conclusion}.}, which is about $10^{7}$ to $10^{10}$ times larger than the neutrino
masses. If neutrinos have sizable scalar/tensor interactions at the
magnitude that concerns the current neutrino scattering experiments,
the large $\nu$MMs may  reach or exceed the known bounds. The connection
between scalar/tensor interactions and large $\nu$MMs has important
implications for future neutrino experiments\,---\,if sizable scalar/tensor
interactions could be found within the sensitivity of future experiments,
then it might imply  large, detectable $\nu$MMs which would motivate
 more elaborate experimental searches, and vice versa.

\section{$\nu$MM from effective interactions\label{sec:basic}}

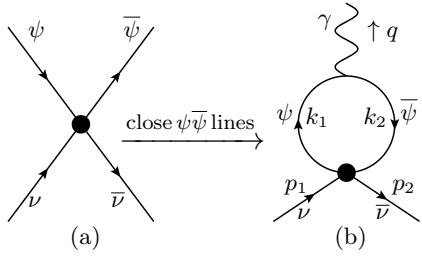
\begin{figure}
\centering

\begin{overpic}[width=5.5cm]
{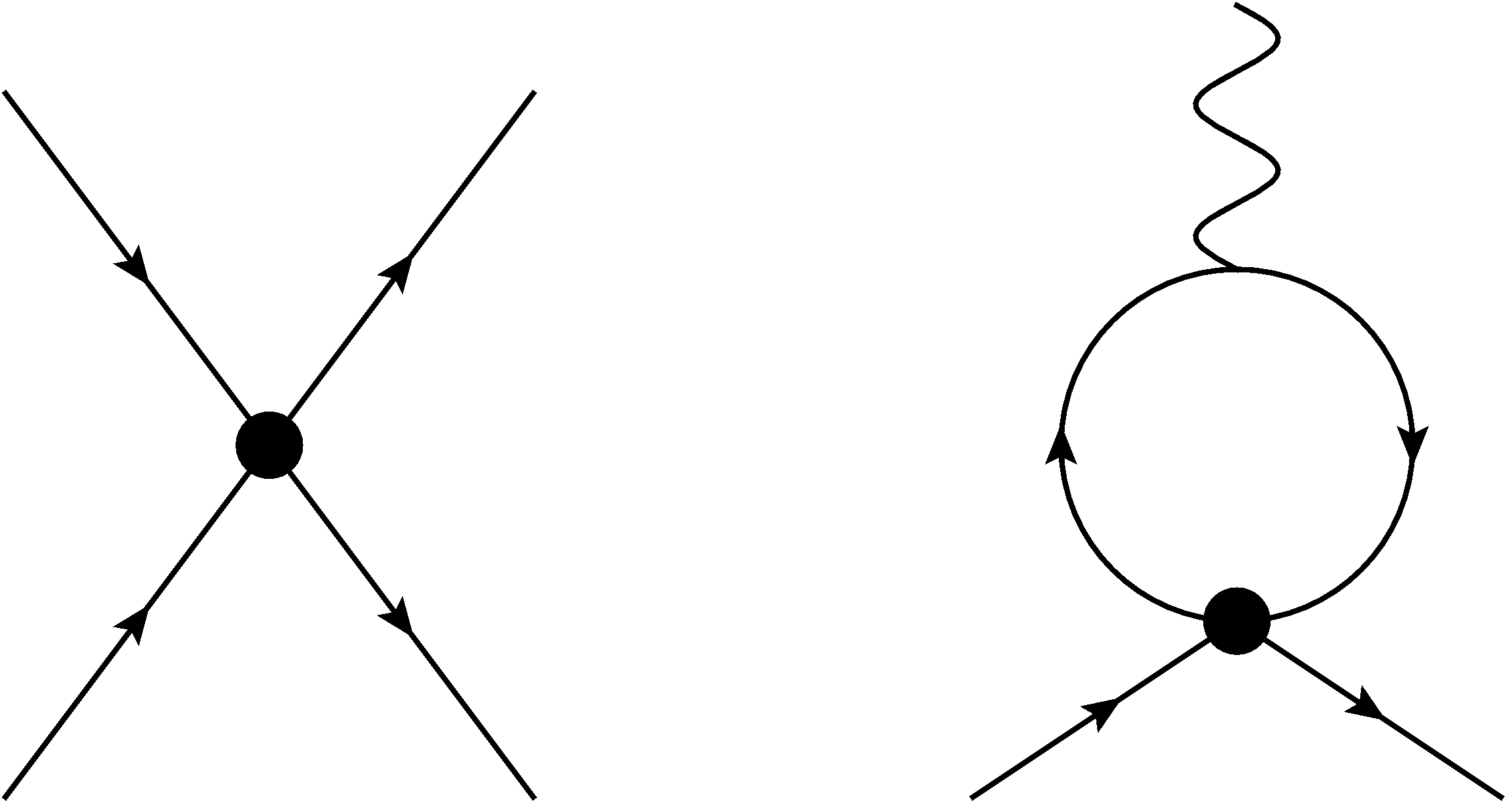} 
\put (5,45) {$\psi$}
\put (28,45) {$\overline{\psi}$} 
\put (5,4) {$\nu$}
\put (25,4) {$\overline{\nu}$} 
\put (65,25) {$\psi$}
\put (95,25) {$\overline{\psi}$} 
\put (70,1) {$\nu$}
\put (89,1) {$\overline{\nu}$} 

\put (15,-5) {(a)}
\put (79,-5) {(b)}

\put (67,8) {$p_1$}
\put (93,8) {$p_2$}
\put (72,24) {$k_1$}
\put (86,24) {$k_2$}

\put (87,45) {$\uparrow q$}
\put (75,48) {$\gamma$}
\put (27,18) {\large $\xrightarrow{{\rm close}\,\psi\overline{\psi}\, {\rm lines}}$}

\end{overpic}

\caption{\label{fig:close_loop}Feynman diagrams showing the connection between
the effective interactions in Eq.~(\ref{eq:spvat}) and $\nu$MMs.}
\end{figure}

In what follows, through  an explicit but short calculation (depicted
in Fig.~\ref{fig:close_loop}), we will show that $\nu$MMs generated
by scalar/tensor interactions are in general proportional to charged
fermion masses instead of neutrino masses. The calculation {\it per se}
will technically explain the reason. To get a deeper insight into
it, after the calculation we will provide an alternative explanation
based on fermion chiralities.

We start by considering the following general effective interactions
of neutrinos\footnote{Here we consider neutrinos in the mass basis and for simplicity, we
focus on one of the three generations. We will discuss the full three-generation
framework in the flavor basis in Sec.~\ref{sec:Conclusion}.

} ($\nu$) and other fermions ($\psi$):
\begin{equation}
{\cal L}\supset G_{X}\left(\overline{\nu}\Gamma\nu\right)\left(\overline{\psi}\Gamma'\psi\right),\label{eq:spvat}
\end{equation}
where $\Gamma$ and $\Gamma'$ can be any Dirac matrices that keep
Eq.~(\ref{eq:spvat}) Lorentz invariant, including 
\begin{equation}
\mathbf{1},\ \gamma_{5},\ \gamma_{\mu},\ \gamma_{\mu}\gamma_{5},\ \sigma_{\mu\nu}\equiv\frac{i}{2}[\gamma_{\mu},\ \gamma_{\nu}],\label{eq:NMM-4}
\end{equation}
and their linear combinations (e.g., $\gamma_{\mu}-\gamma_{\mu}\gamma_{5}$).

In terms of Feynman diagrams, Eq.~(\ref{eq:spvat}) is an effective
vertex of four fermion lines shown in Fig.~\ref{fig:close_loop}\ (a),
relevant to elastic neutrino scattering processes that  are currently
undergoing  precision measurement.  Given such a diagram, one can
close the $\psi$ and $\overline{\psi}$ lines and attach an external
photon line to it, which forms a 1-loop diagram responsible for $\nu$MM
generation. The 1-loop diagram can be evaluated as follows:
\begin{equation}
{\rm Fig.\thinspace\ref{fig:close_loop}\thinspace}({\rm b})=\int\frac{d^{4}k}{(2\pi)^{4}}eG_{X}\overline{u_{2}}(p_{2})\Gamma u_{1}(p_{1})\epsilon^{\mu}(q)\thinspace{\rm tr}_{{\rm loop}},\label{eq:NMM}
\end{equation}
where most notations take the standard convention (e.g., $e$ is the
coupling constant of $\psi$ to the photon, $\epsilon^{\mu}$ is the
photon polarization vector, etc.), all the momenta have been defined
in Fig.~\ref{fig:close_loop} with $k\equiv p_{1}-k_{1}=p_{2}-k_{2}$,
and ${\rm tr}_{{\rm loop}}$ stands for the trace of the loop:
\begin{eqnarray}
{\rm tr}_{{\rm loop}} & = & {\rm tr}\left[\frac{1}{\slashed{k}_{2}-m_{\psi}}\gamma_{\mu}\frac{1}{\slashed{k}_{1}-m_{\psi}}\Gamma'\right]\label{eq:NMM-1}\\
 & = & \frac{{\rm tr}\left[(\slashed{k}_{2}+m_{\psi})\gamma_{\mu}(\slashed{k}_{1}+m_{\psi})\Gamma'\right]}{(k_{2}^{2}-m_{\psi}^{2})(k_{1}^{2}-m_{\psi}^{2})}.\label{eq:NMM-2}
\end{eqnarray}
 Throughout the calculation, we assume neutrinos are Dirac particles
and leave the case of Majorana neutrinos in later discussion.

The trace in Eq.~(\ref{eq:NMM-2}) is crucial to understanding when
the generated $\nu$MM is proportional to $m_{\psi}$. For simplicity,
let us first focus on the scalar interaction ($\Gamma=\Gamma'=\mathbf{1}$),
for which the trace can be easily worked out:
\begin{equation}
{\rm tr}_{{\rm loop}}=m_{\psi}\frac{4\left(k_{1}+k_{2}\right)_{\mu}}{(k_{2}^{2}-m_{\psi}^{2})(k_{1}^{2}-m_{\psi}^{2})}.\label{eq:NMM-5}
\end{equation}
This result can be obtained by noticing that in the numerator of Eq.~(\ref{eq:NMM-2})
only the cross terms  ${\rm tr}[m_{\psi}\gamma_{\mu}\slashed{k}_{1}+\slashed{k}_{2}\gamma_{\mu}m_{\psi}]$
are nonzero. This is because the trace of any product containing an
odd number of $\gamma$ matrices, such as ${\rm tr}\left[\slashed{k}_{2}\gamma_{\mu}\slashed{k}_{1}\right]$
and ${\rm tr}\left[m_{\psi}\gamma_{\mu}m_{\psi}\right]$, must be
zero \cite{peskin}. 

Plugging Eq.~(\ref{eq:NMM-5}) back into Eq.~(\ref{eq:NMM}) and integrating
out $k$, we should have
\begin{equation}
{\rm Eq.\thinspace(\ref{eq:NMM})\thinspace}=m_{\psi}eG_{X}\overline{u_{2}}(p_{2})\left[c_{1}p_{1}^{\mu}+c_{2}p_{2}^{\mu}\right]u_{1}(p_{1})\epsilon_{\mu}(q),\label{eq:NMM-6}
\end{equation}
simply by using the Lorentz invariance. Since the quantity between
$\overline{u_{2}}$ and $u_{1}$ should be both a Dirac scalar and
a Lorentz vector,  we must be able to write it as a linear combination
of $p_{1}^{\mu}$ and $p_{2}^{\mu}$\,---\,here as $c_{1}p_{1}^{\mu}+c_{2}p_{2}^{\mu}$.
Furthermore, since Eq.~(\ref{eq:NMM-5}) is symmetric under $p_{1}\leftrightarrow p_{2}$,
the integral $\int{\rm tr}_{{\rm loop}}d^{4}k$ should lead to a
symmetric result, which implies $c_{1}=c_{2}$. Indeed, this can be
verified by computing the integral manually or using {\tt Package-X} \cite{Patel:2015tea}.
Assuming the effective vertex has the similar UV behavior as the Fermi
effective interaction\footnote{If $G_{X}$ is a constant at arbitrarily high energies, the integral
is divergent. We assume that at low energies $G_{X}$ approximately
remains constant while for $k\rightarrow\infty$, $G_{X}$ decreases
as $k^{-2}$. More specifically, we adopt $G_{X}\propto\frac{1}{k^{2}-m^{2}}$
with $m^{2}\sim G_{X}^{-1}$ standing for the energy scale of this
transition.} and $G_{X}^{-1/2}\gg m_{\psi}\gg m_{\nu}$,  the integral gives
\begin{equation}
c_{1}=c_{2}\approx\frac{i}{8\pi^{2}}\equiv c,\label{eq:NMM-7}
\end{equation}
where ``$\approx$'' means that higher-order terms suppressed by
$m_{\psi}$ and $m_{\nu}$ are not included.

Using the Gordon identity\footnote{See, e.g., Appendix A of Ref.~\cite{Giunti}.}
and Eq.~(\ref{eq:NMM-7}), we can convert Eq.~(\ref{eq:NMM-6}) to
the magnetic moment form 
\begin{equation}
{\rm Fig.\thinspace\ref{fig:close_loop}\thinspace}({\rm b})\approx cm_{\psi}eG_{X}\overline{u_{2}}(p_{2})i\sigma^{\mu\nu}u_{1}(p_{1})q_{\nu}\epsilon_{\mu}(q),\label{eq:nmm-fig-result}
\end{equation}
which implies the following $\nu$MM:
\begin{equation}
\boldsymbol{\mu}_{\nu}\approx\frac{eG_{X}m_{\psi}}{8\pi^{2}},\ ({\rm for\ scalar}).\label{eq:scalar_result}
\end{equation}

As one can see, to get $\boldsymbol{\mu}_{\nu}\propto m_{\psi}$,
the crucial step in the above calculation is that the trace in Eq.~(\ref{eq:NMM-2})
has non-vanishing cross terms (proportional to $m_{\psi}$) while
all the other terms are zero. This is true for $\Gamma=\Gamma'=\mathbf{1}$.
If $(\Gamma,\ \Gamma')=(\gamma_{\nu},\ \gamma^{\nu})$, we would be
 in the opposite situation\,---\,the cross terms become zero while
the other terms are nonzero. A straightforward calculation can confirm
that the $\nu$MM in this case is approximately proportional to $m_{\nu}$
instead of $m_{\psi}$.  

To summarize, whether ${\rm tr}[(m_{\psi}\gamma_{\mu}\slashed{k}_{1}+\slashed{k}_{2}\gamma_{\mu}m_{\psi})\Gamma']$
vanishes or not depends on whether  $\Gamma'$ consists of an odd
or even number of $\gamma$ matrices.  Therefore for the tensor interaction,
we can infer that the result should be proportional to $m_{\psi}$.
Indeed, repeating the previous calculation for $(\Gamma,\ \Gamma')=(\sigma_{\nu\lambda},\ \sigma^{\nu\lambda})$
with the same assumptions gives 
\begin{equation}
\boldsymbol{\mu}_{\nu}\approx\frac{eG_{X}m_{\psi}}{2\pi^{2}}\left[1+\log\left(m_{\psi}^{2}G_{X}\right)\right],\ ({\rm for\ tensor}).\label{eq:tensor_result}
\end{equation}

So far we have technically explained why tensor and scalar interactions
could lead to large $\nu$MMs proportional to $m_{\psi}$. The above
argument based on even/odd numbers of $\gamma$ matrices  can be
more physically interpreted using the concept of chirality flipping. 

First let us examine the chirality of $\nu$MM,
\begin{align}
{\cal L}_{\nu{\rm MM}} & =\boldsymbol{\mu}_{\nu}\overline{\nu}\left[i\sigma^{\mu\nu}q_{\nu}\right]\nu A_{\mu},\label{eq:NMM-8-1}\\
 & =\boldsymbol{\mu}_{\nu}\overline{\nu}\left[i\sigma^{\mu\nu}q_{\nu}\right](P_{L}+P_{R})\nu A_{\mu},\\
 & =\boldsymbol{\mu}_{\nu}\left[\overline{\nu_{R}}\sigma^{\mu\nu}\nu_{L}+\overline{\nu_{L}}\sigma^{\mu\nu}\nu_{R}\right]iA_{\mu}q_{\nu},\label{eq:NMM-8-2}
\end{align}
where $P_{L/R}\equiv\frac{1}{2}\left(\mathbf{1}\mp\gamma_{5}\right)$
and $\nu_{L/R}\equiv P_{L/R}\nu$. Eq.~(\ref{eq:NMM-8-2}) implies
that a $\nu$MM itself has to be chirality flipping, i.e., a left-handed
neutrino, after participating the interaction, will turn into a right-handed
neutrino, and vice versa. 

On the other hand, all vector interactions preserve chirality because
\begin{equation}
\overline{\nu}\gamma^{\mu}\nu=\overline{\nu_{L}}\gamma^{\mu}\nu_{L}+\overline{\nu_{R}}\gamma^{\mu}\nu_{R}.\label{eq:NMM-9}
\end{equation}
So to obtain a nonzero $\nu$MM, we need chirality-flipping sources.
 One of  such sources is a Dirac  neutrino mass term, 
\begin{equation}
m_{\nu}\overline{\nu}\nu=m_{\nu}\left(\overline{\nu_{L}}\nu_{R}+\overline{\nu_{R}}\nu_{L}\right),\label{eq:NMM-10}
\end{equation}
which explicitly shows chirality flipping. In addition, as can be
checked,  tensor or scalar interactions all have the chirality-flipping
property.

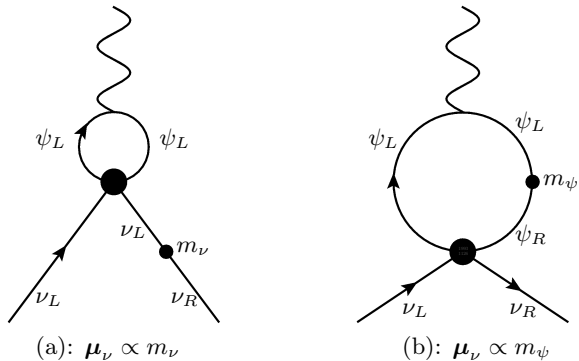
\begin{figure}
\centering

\begin{overpic}[width=7.5cm]
{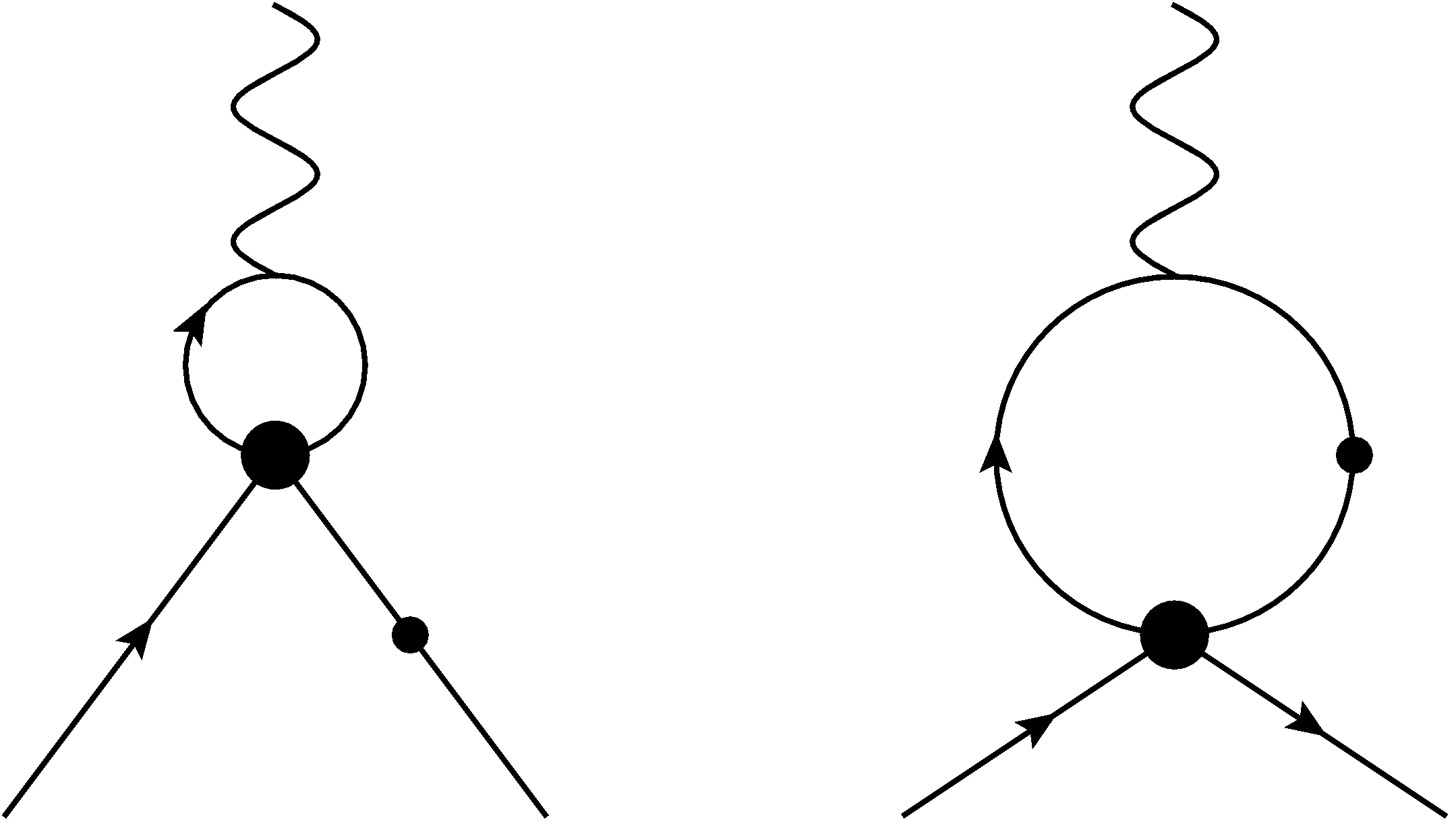} 
\put (5,32) {$\psi_L$}
\put (27,32) {$\psi_L$} 
\put (5,4) {$\nu_L$}
\put (20,16) {$\nu_L$}
\put (30,13) {$m_{\nu}$}
\put (29,4) {$\nu_R$} 

\put (64,32) {$\psi_L$}
\put (90,35) {$\psi_L$} 
\put (95,25) {$m_{\psi}$}
\put (90,15) {$\psi_R$} 
\put (70,2) {$\nu_L$}
\put (89,2) {$\nu_R$} 

\put (5,-5) {(a): $\boldsymbol{\mu}_{\nu}\propto m_{\nu}$}
\put (70,-5) {(b): $\boldsymbol{\mu}_{\nu}\propto m_{\psi}$} 
\put (80,13.5) {\color{very_gray} \scalebox{.2}{1993}}
\put (80,12.5) {\color{very_gray} \scalebox{.2}{1126}}

\end{overpic}
\vspace{0.3cm}

\caption{\label{fig:chirality}Feynman diagrams explaining when $\boldsymbol{\mu}_{\nu}$
are suppressed by the neutrino mass (left) and when by the charged
lepton mass (right). }
\end{figure}

Now let us scrutinize the chirality in the loop diagram. If the 4-fermion
vertex does not flip chirality (e.g., $\Gamma=\gamma_{\nu}$ and $\Gamma'=\gamma^{\nu}$),
then chirality flipping can only be achieved by $m_{\nu}\overline{\nu}\nu$,
as presented in Fig.~\ref{fig:chirality}~(a). It is interpreted
as follows. First, if the left leg is $\nu_{L}$, then the right leg
initially has to be $\nu_{L}$ since the 4-fermion vertex cannot flip
chirality. However, as required by the chirality-flipping property
of $\nu$MM, the right leg eventually should be $\nu_{R}$. So a mass
insertion necessarily appears  on the right leg to achieve the flipping.
In this case, the diagram must be  proportional to $m_{\nu}$.

If the 4-fermion vertex is of tensor or scalar forms {[}see Fig.~\ref{fig:chirality}~(b){]},
then the right leg has the opposite chirality to the left, simply
due to the chirality-flipping property of the vertex.  So we do not
need the mass insertion of $m_{\nu}.$ But we should notice that the
charged fermion also flips its chirality when passing this vertex,
while the photon vertex is not chirality-flipping. To accommodate
both vertices in one loop, a mass insertion of $m_{\psi}$ is necessary,
as marked in Fig.~\ref{fig:chirality}~(b). In this case, the diagram
must be proportional to $m_{\psi}$.

Therefore, we can conclude that if the 4-fermion vertex is chirality-flipping
{\it per se},  then it generates $\boldsymbol{\mu}_{\nu}$$\thinspace\propto\thinspace$$m_{\psi}$;
otherwise it leads to $\boldsymbol{\mu}_{\nu}$$\thinspace\propto\thinspace$$m_{\nu}$.
 This explains why in our previous calculation $\boldsymbol{\mu}_{\nu}$$\thinspace\propto\thinspace$$m_{\psi}$
is obtained for tensor and scalar interactions.

\section{A UV complete example\label{sec:UV}}

The chirality analysis explicates when $\boldsymbol{\mu}_{\nu}$
is proportional to $m_{\psi}$ and when to $m_{\nu}$.  The specific
values of  $\boldsymbol{\mu}_{\nu}$, however, depend on the UV completion
of the effective vertex. Below we would like to study a UV complete
example which introduces a charged scalar $\phi^{\pm}$ interacting
with both left-/right-handed neutrinos ($\nu_{L}$/$\nu_{R}$) and
charged leptons ($\ell_{L}$/$\ell_{R}$):
\begin{equation}
{\cal L}\supset y_{c}\overline{\nu_{L}}\phi^{+}\ell_{R}+y_{s}\overline{\ell_{L}}\phi^{-}\nu_{R}+{\rm h.c.}\label{eq:NMM-11}
\end{equation}
The above terms could originate from left-right symmetric models (LRSM)
\cite{Pati:1974yy,Mohapatra:1974gc,Senjanovic:1975rk}\footnote{Since $\nu_{R}$ appears as an external fermion line in Fig.~\ref{fig:chirality},
the canonical LRSM in which right-handed neutrinos are heavy states
cannot be applied here.} containing  the Yukawa interaction $(\overline{\nu_{L}},\ \overline{\ell_{L}})\Phi(\nu_{R},\ \ell_{R})^{T}$
where $\Phi$ is a bi-doublet, provided that the charged components
in $\Phi$ have generic mass mixing. 

Eq.~(\ref{eq:NMM-11}) can give rise to the 4-fermion effective interactions
of both scalar and tensor forms, if $\phi^{\pm}$ is integrated out:
\begin{equation}
{\cal L}_{{\rm eff}}=\frac{y_{c}y_{s}}{m_{\phi}^{2}}\left(\overline{\nu_{L}}\ell_{R}\right)\left(\overline{\ell_{L}}\nu_{R}\right)+{\rm h.c.},\label{eq:NMM-12}
\end{equation}
which after the Fierz transformation\footnote{To use the   chiral form, see Eqs.~(2.6\,-\,2.7) of Ref.~\cite{Bischer:2018zbd}. }
becomes
\begin{equation}
{\cal L}_{{\rm eff}}=-\frac{y_{c}y_{s}}{8m_{\phi}^{2}}\left(4\overline{\ell_{L}}\ell_{R}\overline{\nu_{L}}\nu_{R}+\overline{\ell_{L}}\sigma^{\mu\nu}\ell_{R}\overline{\nu_{L}}\sigma_{\mu\nu}\nu_{R}\right)+{\rm h.c.}\label{eq:NMM-16}
\end{equation}

Given the Yukawa interactions in Eq.~(\ref{eq:NMM-11}), we know the
specific UV behavior of the effective interactions at high energies.
So $\boldsymbol{\mu}_{\nu}$ can be computed without uncertainties
caused by UV divergences. There are two diagrams responsible for $\boldsymbol{\mu}_{\nu}$: 
\begin{itemize}
\item (i) Fig.~\ref{fig:close_loop}~(b) with the 4-fermion vertex replaced
by a $\phi^{\pm}$ mediator; 
\item (ii) A similar diagram to (i) but the photon is coupled to the $\phi^{\pm}$
mediator.
\end{itemize}
After straightforward loop calculations, the results are: 
\begin{align}
\boldsymbol{\mu}_{\nu}^{{\rm (i)}} & =\frac{em_{\ell}y_{c}y_{s}}{64\pi^{2}m_{\phi}^{2}}\left(3+2\log\frac{m_{\ell}^{2}}{m_{\phi}^{2}}\right),\label{eq:NMM-13}\\
\boldsymbol{\mu}_{\nu}^{{\rm (ii)}} & =-\frac{em_{\ell}y_{c}y_{s}}{64\pi^{2}m_{\phi}^{2}},\label{eq:NMM-14}
\end{align}
corresponding to the contributions of (i) and (ii) respectively. So
in this model the total contribution to the $\nu$MM is 
\begin{equation}
\boldsymbol{\mu}_{\nu}=\boldsymbol{\mu}_{\nu}^{{\rm (i)}}+\boldsymbol{\mu}_{\nu}^{{\rm (ii)}}=\frac{em_{\ell}y_{c}y_{s}}{32\pi^{2}m_{\phi}^{2}}\left(1+\log\frac{m_{\ell}^{2}}{m_{\phi}^{2}}\right).\label{eq:NMM-15}
\end{equation}
This is consistent with our previous discussions based on the effective
operators {[}cf. Eqs.~(\ref{eq:scalar_result}) and (\ref{eq:tensor_result}){]}.
Taking $G_{X}\sim y_{c}y_{s}/(8m_{\phi}^{2})$, we can see that the
effective and the UV complete results agree at the same order of magnitude
while the difference is understandable due to different UV details.

\section{Discussion and Conclusion\label{sec:Conclusion}}

\begin{figure}
\centering

\includegraphics[width=8.8cm]{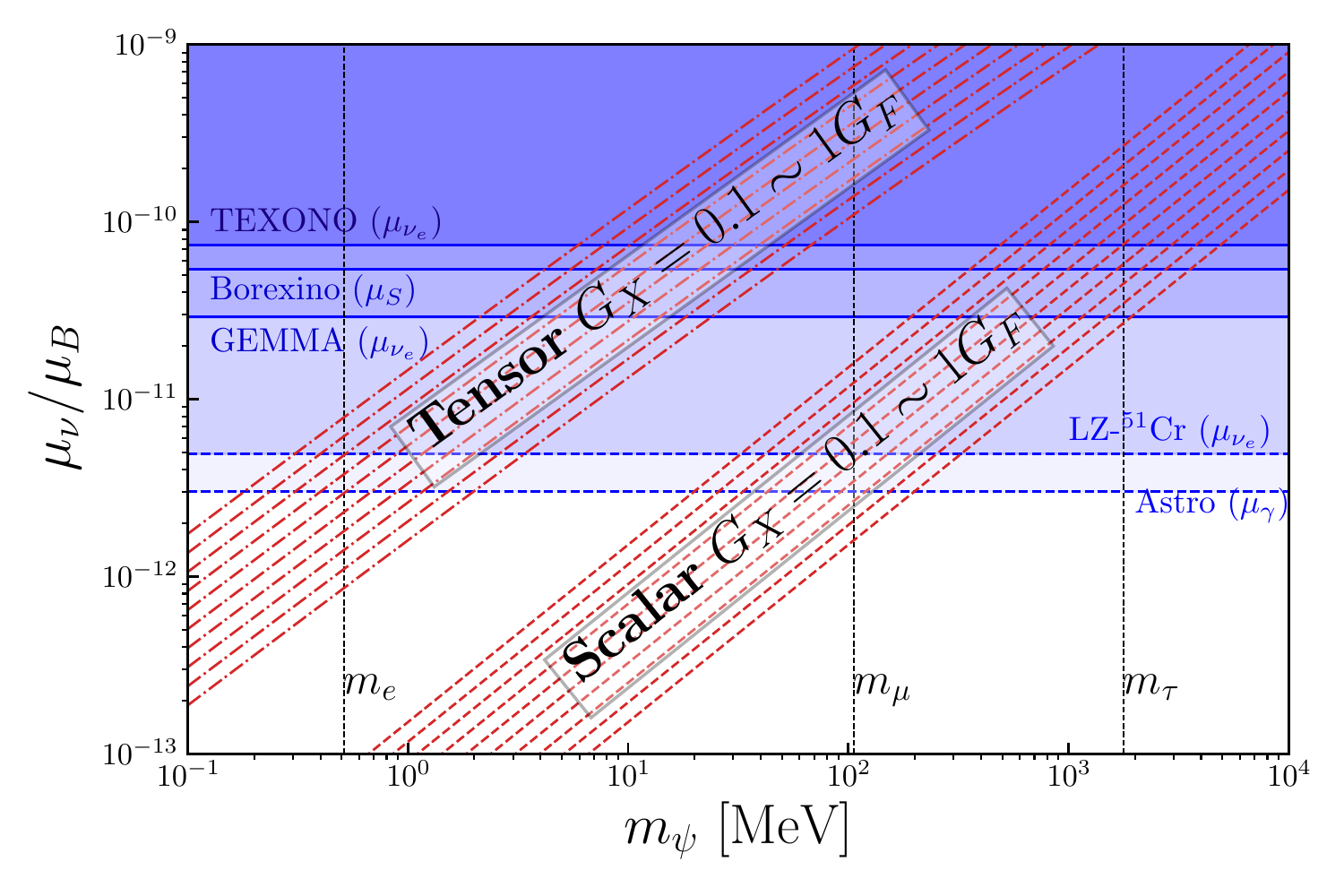}

\vspace{-0.5 cm}

\caption{\label{fig:numeric}$\boldsymbol{\mu}_{\nu}$ predicted by tensor
and scalar interactions (red lines) compared with experimental bounds
(blue). Here the GEMMA, TEXONO and LZ-$^{51}{\rm Cr}$ bounds are
only for the effective magnetic moment $\boldsymbol{\mu}_{\nu_{e}}$
defined in Eq.~(\ref{eq:NMM-18}); the Borexino bound is based on
solar neutrinos which should be applied to $\boldsymbol{\mu}_{S}$
given in Eq.~(\ref{eq:NMM-19}); and the astrophysical (Astro) bound
should be applied to $\boldsymbol{\mu}_{\gamma}$ in Eq.~(\ref{eq:NMM-20}).}
\end{figure}

Throughout the paper we have only considered the case of Dirac neutrinos.
For Majorana neutrinos, our conclusions would be similar but need
slight modification. As is well known,  Majorana neutrinos can only
have transition magnetic moments, meaning that the  corresponding
term $\overline{\nu}_{i}\sigma^{\mu\nu}\nu_{j}q_{\nu}A_{\mu}$ may
exist only if $i\neq j$ ($i,\ j=1,\ 2,\ 3$ denote the  mass eigenstates
of neutrinos; $\nu\equiv\nu_{L}+\nu_{L}^{c}$ is a Majorana spinor
so that $\nu=\nu^{c}$). Viewed from fermionic degrees of freedom,
the transition from $\nu_{i}\rightarrow\nu_{j}$ is essentially equivalent
to the aforementioned chirality flipping as the initial and final
neutrinos are two different Weyl spinors.  Therefore, for Majorana
neutrinos we simply need the replacement $(\nu_{R},\ \nu_{L})\rightarrow(\nu_{Li}^{c},\ \nu_{Lj})$
in the above analyses.

The analyses in this paper can be readily extended to include three
flavors. First, Eq.~(\ref{eq:spvat}) can be modified to the flavor-dependent
form:
\begin{equation}
{\cal L}\supset G_{X}^{\alpha\beta}\left(\overline{\nu_{\alpha}}\Gamma\nu_{\beta}\right)\left(\overline{\psi}\Gamma'\psi\right),\label{eq:NMM-17}
\end{equation}
where $\alpha$, $\beta=e$, $\mu$, $\tau$ are flavor indices. Then
since we know that for tensor and scalar interactions neutrino masses
make negligible contributions to $\nu$MMs, neutrinos can be treated
as massless particles in the calculation, which would lead to flavor-dependent
$\boldsymbol{\mu}_{\nu}^{\alpha\beta}$ in Eqs.~(\ref{eq:scalar_result})
and (\ref{eq:tensor_result}) with only $G_{X}$  replaced by $G_{X}^{\alpha\beta}$.
Note that many experimental measurements actually produce constraints
on combinations of some $\boldsymbol{\mu}_{\nu}^{\alpha\beta}$. For
example, $\nu_{e}$-$e$ scattering experiments with negligible baselines
are sensitive to  the effective magnetic moment of $\nu_{e}$ below
\cite{Kouzakov:2017hbc}:
\begin{equation}
\boldsymbol{\mu}_{\nu_{e}}^{2}=\sum_{\beta}\left|\boldsymbol{\mu}_{\nu}^{e\beta}\right|^{2}.\label{eq:NMM-18}
\end{equation}
For solar neutrino experiments, the effective magnetic moment being
constrained is ~\cite{Giunti:2014ixa} 
\begin{equation}
\boldsymbol{\mu}_{S}^{2}=\sum_{j,k=1}^{3}\left|U_{ek}^{M}\right|^{2}\left|\boldsymbol{\mu}_{\nu}^{jk}\right|^{2},\label{eq:NMM-19}
\end{equation}
where $U_{ek}^{M}$ is the effective neutrino mixing with the matter
effect included, and $\boldsymbol{\mu}_{\nu}^{jk}$ is the mass-basis
form of $\boldsymbol{\mu}_{\nu}^{\alpha\beta}$. In addition, for
plasmon decay ($\gamma^{*}\rightarrow\overline{\nu}\nu$) \cite{Raffelt:1999gv},
one can define the following effective magnetic moment, 

\begin{equation}
\boldsymbol{\mu}_{\gamma}^{2}=\sum_{j,k=1}^{3}\left|\boldsymbol{\mu}_{\nu}^{jk}\right|^{2}=\sum_{\alpha,\beta}\left|\boldsymbol{\mu}_{\nu}^{\alpha\beta}\right|^{2},\label{eq:NMM-20}
\end{equation}
which is useful in interpreting the astrophysical bounds.

The values of $\boldsymbol{\mu}_{\nu}$ given by Eqs.~(\ref{eq:scalar_result})
and (\ref{eq:tensor_result}) depend on the UV completion of the effective
vertices. Being model dependent implies that $\boldsymbol{\mu}_{\nu}$
could be much smaller or larger than Eqs.~(\ref{eq:scalar_result})
and (\ref{eq:tensor_result}) in  particular models. For example,
if it is UV completed by a neutral scalar $\phi$ with Yukawa interactions
$\overline{\nu}\nu\phi$ and $\overline{\psi}\psi\phi$, then the
loop diagram naively gives zero $\boldsymbol{\mu}_{\nu}$. However,
since it breaks $SU(2)_{L}$, usually this model is a fragment of
some more complete gauge invariant models, in which $\phi$ would
be the neutral component of a Higgs multiplet and be accompanied
with charged scalar bosons. The full calculation, including contributions
from charged bosons, may again lead to large nonzero $\boldsymbol{\mu}_{\nu}$.

Taking Eqs.~(\ref{eq:scalar_result}) and (\ref{eq:tensor_result})
as the typical values of $\boldsymbol{\mu}_{\nu}$ generated by the
effective tensor and scalar interactions, we plot them in Fig.~\ref{fig:numeric}
together with terrestrial (TEXONO \cite{Wong:2006nx}, Borexino
\cite{Arpesella:2008mt}, GEMMA \cite{Beda:2012zz}, and LZ equipped
with an intensive $^{51}{\rm Cr}$ radiative source \cite{Coloma:2014hka})
and astrophysical \cite{Raffelt:1999gv} bounds. Currently the effective
coupling $G_{X}$ can be constrained by various elastic neutrino scattering
data from CHARM II, LSND, TEXONO, Borexino, COHERENT, etc. In general,
these experiments have $G_{X}$ sensitivity ranging from $0.1$ to
$1\thinspace G_{F}$ \cite{Sevda:2016otj,Rodejohann:2017vup,Lindner:2018kjo,AristizabalSierra:2018eqm},
depending on the neutrino flavors, the charged fermion $\psi$, the
specific forms of new interactions, etc. With these details involved
and the uncertainties of theoretical predictions due to the UV incompleteness,
here we refrain from more specific discussions and show merely two
bands (red) of $G_{X}=0.1\sim1G_{F}$ in Fig.~\ref{fig:numeric}.
In the future, the DUNE near detector and some reactor-based coherent
neutrino scattering experiments may significantly improve the sensitivity
by one or two orders of magnitude \cite{Bischer:2018zcz,Lindner:2016wff}.

The significance of Fig.~\ref{fig:numeric} showing the red bands
and the blue limits in the same windows is manifold. If, e.g., $G_{X}=0.1G_{F}$
for tensor interactions had been probed in neutrino-electron scattering
experiments, it would imply a large $\nu$MM ($\boldsymbol{\mu}_{\nu}\sim10^{-12}\mu_{B}$)
that could be observed by improving $\nu$MM experiments by one order
of magnitude. In addition, since the same coupling strength for $\psi=\mu$
and $\tau$ would lead to too large $\boldsymbol{\mu}_{\nu}$, it
would imply that in model building, $G_{X}$ for these two flavors
must be suppressed, which is of theoretical importance. On the other
hand, if in the future we reach much more solid and stringent bounds
on $\boldsymbol{\mu}_{\nu}$ (currently LZ-$^{51}{\rm Cr}$ is only
a proposal and the astrophysical bound could be altered in non-standard
scenarios), it will disprove the presence of sizable tensor and scalar
interactions, which is still of importance for both experimental searches
and theoretical model building.

The last comment concerns neutrino masses. It has been commonly discussed
 in the literature (reviewed in Ref.~\cite{Giunti:2014ixa}) that
the new physics leading to large $\nu$MMs  usually generates too
large  neutrino masses. This can be understood by simply noticing
that in the absence of chirality-flipping interactions the generated
$\nu$MM is proportional to $m_{\nu}$. There have been various approaches,
however, to get a large  $\nu$MM while keeping $m_{\nu}$ small.
One possibility is to avoid it from being proportional to $m_{\nu}$,
which has been discussed in Refs.~\cite{Shrock:1974nd,Kim:1976gk,Marciano:1977wx,Beg:1977xz,Shrock:1982sc,Duncan:1987ki,Liu:1987nf,Rajpoot:1990hj,Czakon:1998rf,Nemevsek:2012iq,Boyarkin:2014oza}.
 For example, in the left-right symmetric model with Dirac neutrinos,
$\boldsymbol{\mu}_{\nu}\propto m_{\ell}$ can be obtained (see, e.g.,
Eq.~(2.29) in \cite{Shrock:1982sc}) via  the charged current (CC)
interaction $\overline{\ell}_{L}\gamma^{\mu}\nu_{L}W_{L\mu}^{-}$
and its right-handed partner $\overline{\ell}_{R}\gamma^{\mu}\nu_{R}W_{R\mu}^{-}$
where $W_{L}^{\pm}$ and $W_{R}^{\pm}$ are the charged gauge bosons
of $SU(2)_{L}$ and $SU(2)_{R}$ with small mass mixing.  From the
point of view of effective interactions adopted in this paper, it
is straightforward to understand the result. The left- and right-handed
CC interactions with mixing can give rise to the effective interaction
$(\overline{\ell}_{L}\gamma^{\mu}\nu_{L})(\overline{\nu}_{R}\gamma^{\mu}\ell_{R})$,
which after the Fierz transformations becomes a chirality-flipping
scalar interaction $2(\overline{\nu}_{R}\nu_{L})(\overline{\ell}_{L}\ell_{R})$.
This should lead to $\boldsymbol{\mu}_{\nu}\propto m_{\ell}$ according
to our conclusion on such scalar interactions. Therefore, the calculation
in the previous studies confirms our conclusion on the effective interactions.
In addition to this model, there are various other models proposed
for large $\nu$MMs \cite{Voloshin:1987qy,Alles:1989tr,Barr:1990um,Barr:1990dm,Babu:1990hu,Deshpande:1991ih,Babu:1992vq,Chang:1992fs,Frank:2000na,Lindner:2017uvt}.
Although building models for large $\nu$MMs is not the focus of this
paper, our conclusion indicates that one may preferably introduce
chirality-flipping interactions to obtain large $\nu$MMs  because
in this situation, $\boldsymbol{\mu}_{\nu}$ is  proportional to $m_{\psi}$
instead of $m_{\nu}$, and the generation of $\nu$MMs can be detached
from the generation of neutrino masses.

In conclusion, our analysis reveals that large $\nu$MMs may be potentially
related to sizable tensor and scalar interactions, and vice versa.
The experimental and theoretical significance of the interplay will
be explored in further studies.

\begin{acknowledgments}
X.J.X would like to thank Evgeny Akhmedov and Alexei Smirnov for
many helpful conversations on $\nu$MMs, Robert Shrock for discussions
on $\nu$MMs in the left-right symmetric model, and especially Rabindra Mohapatra
for insightful discussions on our previous work \cite{Rodejohann:2017vup}
which gradually developed into the initial idea of this work.
\end{acknowledgments}

\bibliographystyle{apsrev4-1}
\bibliography{ref}

\end{document}